# A Theoretical Model For Artificial Learning, Memory Management And Decision Making System

Ravin Kumar

*Department of Computer Science*

*Meerut Institute Of Engineering And Technology, meerut-250005, INDIA.*

Email id: mr.ravin_kumar@hotmail.com

*Abstract*- **Human beings are considered as the most intelligent species on Earth. The ability to think, to create, to innovate, are the key elements which make humans superior over other existing species on Earth. Machines lack all those elements, although machines are faster than human in aspects like- computing, equating etc. But humans are still more valuable than machines, due to all those previously discussed elements. Various models have been developed in last few years to create models that can think like human beings, but are not completely successful. This paper presents a new theoretical system for learning, memory management and decision making that can be used to develop highly complex systems, and shows the potential to be used for development of systems that can be used to provide the essential features to the machines to act like human beings.**

## I. INTRODUCTION

The ability to think have made humans more important than other species, this have enabled us for development of science, technology, and our own civilization. Many attempts have been made to artificially create a system that can show all those features that made humans more important than any other species, features like- thinking capability, feelings, emotions etc. Some models have been made using inspiration from nature, some purely from mathematics, and are still in use today. Some of these structures are given below:

**Artificial Neural Network**- Artificial neural networks learn to compute a function using example inputs and outputs. Neural networks have been used for a variety of applications, including pattern recognition, classification [1], and image understanding [2].

**Fuzzy Logic**- This approach uses the model based on degree of truth rather than just exact true or exact false states. It is the from many-valued logic in which the truth values of variables may be any real number between 0 and 1 [3].

**Genetic Algorithm**- This approach is inspired by the process of natural selection , these are generally used to solve optimization and searching problems by relying on bio-inspired operations such as mutation [4] and cross over [5].

Learning is improvement on performance with experience over time. There are mainly three types of learning mechanism that have been developed artificially by human beings, and these are given below as:

**Supervised Learning**- In this type of learning, a tutor is present to guide the mahine by providing its feedback. In this approach the data is labeled [6-7].

**Unsupervised Learning**- In this type of learning, no tutor is present and machine learn by trying on its own. In this approach the data is not labeled [8-9].

**Reinforcement Learning**- In this type of learning, the main concern is how software agents ought to take *actions* in an *environment* so as to maximize some notion of cumulative *reward* [10].

All these above discussed models are used for solving problem using machine learning, and have shown some effective results also, But they all still are unable to make a system which is as much powerful as our human brain. This paper presents a theoretical system which when implemanted can be used for designing of a complex machine such as human brain, which will be capable of learning, managing memory and taking decisions effectively. The organisation of this paper is as follows:

Section 2 presents the description of proposed system, Section 3 present the learning, memory management and decision making process in details. Results and Discussion are in Section 4. Conclusions are given in Section 5, and references are given at the last.

## II. PROPOSED SYSTEM

In the proposed system, Memory Pixel is the proposed to be the smallest unit in which a individual data can be represented.

**Memory Pixel** consist of four fields namely- Color, Intensity, Device_id and Data. It can be represented as:

| Color | Intensity | Device_id | Data |
|-------|-----------|-----------|------|

Fig. 1 Representation of Memory Pixel

Memory pixel consist of following fields:

i)**Color**: It represents the relatedness of data. Color represents feelings, actions, emotions etc.

ii) **Intensity**: This field represents the importance of data. Higher value represents more important data.

ii) **Device_id**: It represents that peripheral device, from where either the data have been received, or the command have to be send to perform some action.

iv) **Data**: It contains the data, which is either obtained from the peripheral device, or are generated during the processing.

**Memory screen** is the collection of memory pixels collected at the same time 't', and other than that, it also have three fields- Color, Sequence_number and Data. Here Color represents the color field, which is in majority, in the memory pixel of that particular memory screen. Sequence_number represents the sequence number of the screen pixel. Data represents the most commonly occuring data, in the memory pixels of the memory screen.

The overall architectural design of the proposed system is given below:

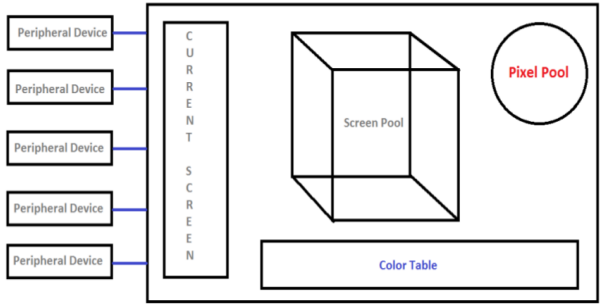

Fig. 2 Decision Making System.

**Peripheral Device**, it acts as the artificial sense organ which provide data to the Decision Making System, and takes the command from the system, to perform the action which is decided by the system, each device have a unique id (*i.e. device_id* ).It sends data in the form of Memory Pixel, each having the device_id of their respective peripheral device.

**Current Screen**, it holds the current data that is received from all the connected peripheral devices, in the form of Memory Screen, and generates the memory screen, and assign it the sequence number (in ascending order).

**Screen Pool**, it holds all those previously received memory screens, which contain important data ( *importance is decided by the intensity of the memory screens, along with its color field* ). It can be represented as:

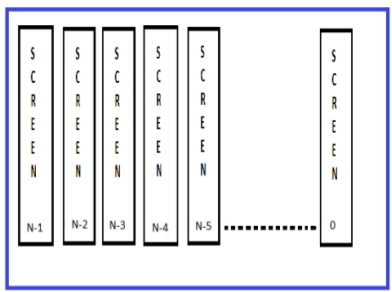

Fig. 3 Screen Pool containing 'N' memory screens.

**Pixel Pool**, It is the place where all free memory pixels are placed. When the data is received from the peripheral device then these memory pixel moves to current screen, receives the data and form a memory screen for the screen pool. And when a memory screen is removed from the screen pool, then the contained memory pixels are moved back to the pixel pool.

**Color Table**, It contains three attributes namely- Color, Memory Screen Number, and Data. The color field represents the color that is available in the memory pool and memory pixel. Memory screen number represents the sequence number of that memory screen which have same color value, and the data field contains the value of the data field of that memory screen.

Fig. 4 Color Table

There is a continuous process, which helps in removing the irrelevent data from the Screen Pool, and updates the attribute values of the Color Table. This process decrement the value of Intensity of the memory pixels, that are present in various memory screens, except those in the r*oot memory screens (* Root memory screen contains those functions, emotions and their associated data, that can not be over written *)*. In this process, after every 'w' seconds ( 'w' can be assigned any value greater than 0 ) the intensity of all memory pixel is decremented by 1.When the value of the intensity becomes -1, then that particular memory pixel get removed from the memory screen, and send back to the pixel pool. This process is applied to all the memory screens except the root memory screens. And the memory screens that have their sequence number in the Color Table, their memory pixels intensity is incremented by a '**r**' (where, $0 < r < 1$ ).

When their is no memory pixel present in the memory screen, then that particular memory screen is removed from the screen pool, and all its entries in the Color Table are also removed.

III. MEMORY MANAGEMENT AND DECISION MAKING SYSTEM

The learning, memory management and decision making processes of the proposed system is discussed below. Firstly, memory management process is discussed, and then the decision making process is discussed, learning mechanism continues in both these steps.

**Memory Management Procedure :**

It is a continuous process, which is applied after every 'p' seconds.The parameter p can be set any value that is above 0, but for better performance p should be greater than 0 and less than 1.

**Step1**: Take inputs from various peripheral devices, and store that data into the Data field of the memory pixels ( memory pixels are received from the pixel pool ), and assign the peripheral device number to the device_id field of their respective memory pixels.

**Step2**: All these memory pixels are then collected in the current screen, to form a new memory screen. Sequence number of each memory screen is assigned in the increasing order.

**Step3**: In the current screen, data field value of each memory pixel is checked with the Data field value of the Color Table, and If the Data values matches then, the corresponding color value is assigned to that memory pixel. And if values do not matches, then new Color value is assigned to the memory pixel and maximum intensity is assigned to that memory pixel, and a new entry is made in the color table, having same color value as that of the memory pixel.

**Step4**: If the color value was previously present in the color table then, the intensity of memory pixel is set to 'I', (where I < Maximum Intensity). And if the color value is newly assigned to the Color Table then, the intensity of memory pixel is set to maximum.

**Step5**: Now in the current screen, majority of the color is checked, and the color which is used by majority of pixels, is assigned to the Color field of the memory screen ( which is created by current screen). Now each field of the memory pixel, and memory screen (created by current screen) have a value. The memory screen is then send to the screen pool.

**Step6:** The continuous process helps in updating the Color Table, removing their irrelevant information, and free those memory pixels that have data of no importance.

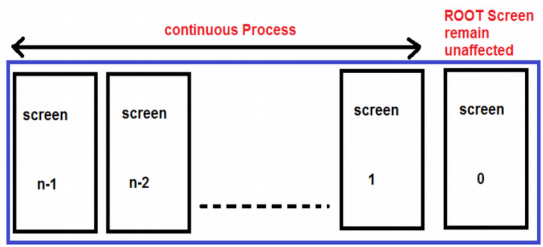

Fig. 5 Screen Pool containing 'N' memory screens.

**Step7**: To receive new data from outside, GOTO Step 1.

**Decision Making Procedure :**

It is also a continuous process, which help in process like - learning, decision making etc.

**Step1**: During the memory management process, when all the field values are assigned to the current screen, then the color value and the data field value of the newly created memory screen ( present in the current screen), is matched with the values of the Color Table.

**Step2**: IF all the value matches, then the system go to that memory screen, whose sequence number was present along with the matching data in the Color Table. And perform the action described in the data field of that particular memory screen.

**Step3**: IF only the color value matches, then all the memory pixels of that particular color, that are present in the memory screen, whose sequence number is present in the Color Table, having the same color value, as that of the newly created memory screen, are copied to the newly created memory screen.

**Step4**: IF Color field do not matches, then the next major color is calculated, and all the searching is performed on the basis of that color. And if color field of that next major color is matched with the color field of Color Table, then search for next major color, and so on, till either a match is found, or all the color have been checked.

**Step5**: IF their is no match found in the step 4, then no action is performed.

IV. RESULT AND DISCUSSION

The proposed theoretical system provides a new approach for learning mechanism, by introducing the memory pixels, relating data with previous knowledge becomes more easy. With the introduction of screen pool the unnecessary memory usage is avoided, this feature is inspired by the concept of "forgetting things" in human beings. Using real time data, it is possible to take on the moment decisions with the help of color tables. More than one screen pools can be used for better performance, and more than one decision making systems can be combined for achieving better results. Depending upon the type of machine we want to design we can add more fields in the memory pixel, and the color table.

V. CONCLUSION

The proposed theoretical system can be used for developing new effective machines that can provide much more capabilities and can work with highly complex data. This theoretical system can be used for developing much complex structures such as human brain. With the introduction of this system more research work can be done in field and more effective and efficient systems can be build.

REFERENCES

[1] Daniel A. Jim enez Calvin Lin, "Dynamic Branch Prediction with Perceptrons", The University of Texas at Austin, TX 78712.
[2] D. A. Jim´enez and N. Walsh. Dynamically weighted ensem-ble neural networks for classification. InProceedings of the1998 International Joint Conference on Neural Networks, May1998.
[3] C. C. Lee. ―Fuzzy logic in control systems: Fuzzy logic controller-part 1,‖ IEEE trans. Syst., Man, Cybern., Vol. 20, pp. 404-418, 1990.
[4] John J. Grefenstette, "Optimization of control parameters for genetic algorithms", IEEE Transaction on Systems, Man, Cyber-netics, SMC-16(1):122-128,1986.
[5] Eiben, A. E. et al (1994). "Genetic algorithms with multi-parent recombination". PPSN III: Proceedings of the International Conference on Evolutionary Computation. The Third Conference on Parallel Problem Solving from Nature: 78–87. ISBN 3-540-58484-6.
[6] S. B. Kotsiantis, "Supervised Machine Learning: A Review of Classification Techniques", Informatica (2007) 249-268.
[7] Roy, A. (2000), On connectionism, rule extraction, and brain-like learning. IEEE Transactions on Fuzzy Systems, 8(2): 222-227.
[8] D. J. C. MacKay. Information Theory, Inference, and Learning Algorithms. Cambridge UniversityPress, 2003.
[9] A. Beygelzimer, S. Kakade, and J. Langford. Cover trees for nearest neighbor. In Proceedings of 23rd International Conference on Machine Learning (ICML 2006), 2006.
[10]Zhang,W.,&Dietterich,T.G.(1995).A reinforcement learning approach to job-shop scheduling. In Proceedings of the International Joint Conference on Articial Intelligence.

BIBLIOGRAPHY

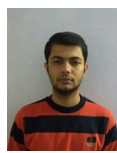

Ravin Kumar is doing his Bachelor of Technology(2013-2017 ) in Computer Science and Engineering, from Meerut Institute Of Engineering and Technology, meerut, U.P, INDIA. During his graduation, he have also done work in other related fields including - a new CPU scheduling algorithm, modified counting sort algorithm, and a computer ransom ware.